\documentclass[aps,pra,10pt,twocolumn,showpacs,floatfix,superscriptaddress]{revtex4-1}
\usepackage{fancyhdr,amsmath,amssymb,amsfonts,bbm,graphicx,hyperref,times,indentfirst,lipsum,mwe,color}
\usepackage{bbold}

\usepackage{natbib}
\usepackage{hyperref}
\hypersetup{
colorlinks=true,
linkcolor=blue,
citecolor=blue,
filecolor=magenta,
urlcolor=cyan
}

\usepackage[capitalize]{cleveref}

\def\Tr{\hbox{Tr}}

\newcommand{\be}{\begin{equation}}
\newcommand{\ee}{\end{equation}}
\newcommand{\bae}{\begin{eqnarray}} \newcommand{\eae}{\end{eqnarray}}

\def\EE{\mathbbm{E}}

\def\Tr{\hbox{Tr}}

\def\sigmaCM{\boldsymbol{\sigma}}

\begin{document}
\title{Unconditional mechanical squeezing via back-action evading measurements and non-optimal feedback control}
\author{Antonio Di Giovanni}
\affiliation{Dipartimento di Fisica ``Aldo Pontremoli'', Universit\`a degli Studi di Milano, I-20133 Milano, Italy}
\author{Matteo Brunelli}
\affiliation{Cavendish Laboratory, University of Cambridge, Cambridge CB3 0HE, United Kingdom}
\author{Marco G. Genoni}
\email{marco.genoni@fisica.unimi.it}
\affiliation{Dipartimento di Fisica ``Aldo Pontremoli'', Universit\`a degli Studi di Milano, I-20133 Milano, Italy}
\affiliation{INFN - Sezione di Milano, I-20133 Milano, Italy}
\begin{abstract}
Backaction-evading (BAE) measurements of a mechanical resonator, by continuously monitoring a single quadrature of motion, can achieve precision below the zero-point uncertainty. When this happens, the measurement leaves the resonator in a quantum squeezed state. The squeezed state so generated is however conditional on the measurement outcomes, while for most applications it is desirable to have a deterministic, i.e., unconditional, squeezed state with the desired properties. In this work we apply feedback control to achieve deterministic manipulation of mechanical squeezing in an optomechanical system subject to a continuous BAE measurement. We study in details two strategies, direct (Markovian) and state-based (Bayesian) feedback. We show that both are capable to achieve optimal performances, i.e., a vanishing noise added by the feedback loop. Moreover, even when the feedback is restricted to be a time-varying mechanical force (experimentally friendly scenario) and an imperfect BAE regime is considered,
the ensuing non-optimal feedback may still obtain significant amount of squeezing. In particular, we show that Bayesian feedback control is nearly optimal for a wide range of sideband resolution. 
Our analysis is of direct relevance for ultra-sensitive measurements and quantum state engineering in state-of-the-art optomechanical devices.
\end{abstract}
\maketitle
\section{Introduction}\label{s:intro}
The accuracy with which the position of an oscillator continuously monitored in time can be resolved has a fundamental limit, known as the standard quantum limit (SQL)~\cite{QuantumNoiseRev, WeakMeasRev, OnofrioRev}.   
Backaction-evading (BAE) measurements have been proposed as a way to circumvent this limit by restricting the measurement to a single quadrature of motion \cite{Braginsky1, QNDThorne78, WeakMeasRev}. BAE measurements can be thought as \emph{classical} measurements embedded in a quantum framework. Classical in the sense that they allow repeated measurements with arbitrary precision, since no backaction (stemming from the non-commutative nature of the observables) corrupts them. At the same time, evading  this constraint is connected with the emergence of quantum properties of the object being measured~\cite{WeakMeasRev}. 

Cavity optomechanics affords an extremely effective way to control and monitor mechanical motion~\cite{OptoRev,BowenMilburnBook}. A simple way to implement a BAE measurement of a mechanical quadrature is to drive an optomechanical cavity on both mechanical sidebands~\cite{Braginsky1, CMJ2008}. This is referred as two-tone BAE scheme and has been demonstrated both at microwave~\cite{BAEOpto09, BAEOpto14} and optical~\cite{Shomroni19} frequencies, with sensitivities approaching the SQL. 
BAE measurements play a fundamental role for ultra-sensitive force measurements in large-scale interferometers as well as in tabletop experiments. They are key for reconstructing mechanical motion, e.g. for probing independently prepared mechanical squeezed states~\cite{Lecocq15,Lei16}. They have also been studied and demonstrated in atomic ensembles~\cite{BAESpin15,Wasilewski10,HybridBAE}, extended to different regimes, e.g. stroboscopic~\cite{MeStrobo20}, and to collective quadratures of two resonators~\cite{TsangCohCanc, TsangQND,TwoModeBAE,TwoModeBAEExp,HammererEPR, BAEBEC,HybridBAE}.

Achieving uncertainties below the SQL is tied to the appearance of quantum squeezing, whereby fluctuations along the measured quadrature are smaller than the zero-point level~\cite{Wiseman93, QuantumMeasBook, JacobsSteckRev}. 
Squeezing is a well-known resource for continuous-variable quantum information~\cite{GaussRevFOP,GaussRev,Gottesman2001,Menicucci2006,Oussama15,Oussama18} and for quantum metrology~\cite{Kwon19,Garbe19,Gessner2020}, in particular for the estimation of Hamiltonian and environmental parameters~\cite{Monras2006,Monras2007,Gaiba2009,Genoni2011,Genoni2013,Carrara2020}, with the paradigmatic example of gravitational-wave detection~\cite{Caves1981,AdvancedLigo2019,Virgo2019}. Therefore, besides granting precision measurements, optomechanical BAE measurements of a single motional quadrature also provide an effective tool for \emph{quantum state preparation} of mechanical squeezed states. However, quantum states prepared via this strategy, or in general via time-continuous monitoring, are conditional on the stream of measurement outcomes~\cite{Genoni2013feedback,Genoni15,MeBAE19}, which makes them less practical for real-time manipulation. This issue can be remedied by implementing a feedback loop to render squeezed states unconditional, i.e. measurement independent~\cite{QuantumMeasBook}. In this way, feedback control can achieve deterministic manipulation of mechanical squeezing. 
 
So far, feedback control of mechanical motion have mainly focused, both theoretically and experimentally, on cooling~\cite{Mancini1998,Doherty99,Hopkins2003,Genes2008,Doherty2012PNAS,Hofer15,Wilson2015,Shudir2017,Rossi2017,Rossi2018},
which is a prerequisite for implementing most quantum protocols. In particular, feedback cooling to the ground state (residual thermal occupancy $\bar n=0.29$) was recently demonstrated in soft-clamped membranes~\cite{Rossi2018} and cooling to microkelvin temperatures ($\bar n=4$) by feedback only was reported in optically levitated nanoparticles~\cite{Tebbenjohanns20}. Given this tremendous success, extending feedback control to quantum properties of mechanical motion, such as squeezing, seems the next logical step and within experimental reach. However, control protocols for mechanical squeezing are much less explored.  
To date, proposals in this direction have focused on obtaining mechanical squeezing via monitoring and state-based feedback~\cite{Rusko2005,CMJ2008}, designing alternative feedback protocols based on ancillary two-level systems~\cite{Genoni2015} or exploiting parametric amplification~\cite{Szorkovszky11,Vinante2013,Pontin2014} or via open-loop control protocols~\cite{Asjad2014}.

In this work we study in details two feedback strategies, direct (Markovian) feedback and state-based (Bayesian) feedback, in combination with time-continuous BAE measurements, to obtain unconditionally mechanical squeezing. We show that both strategies are capable to achieve optimal performances, i.e., vanishing noise added by the feedback loop, in suitable conditions. However, the range of parameters where this occurs greatly differs, highlighting crucial differences between the two approaches. In both cases, we first determine the ideal feedback loop, i.e., the one that adds no noise, to be implemented in a perfect BAE regime. For this case we obtain simple analytical expressions, which we then use as a benchmark to evaluate the effects of introducing physical limitations and non-idealities. In particular, the main sources of limitations we considered are (i) the case in which the feedback is restricted to be a time-varying mechanical force and (ii) imperfect BAE regime where counter-rotating terms cannot be neglected. 
Remarkably, even when assuming both restrictions, we show that Bayesian feedback is nearly optimal (vanishing added noise) across several orders of magnitude of the sideband parameter $\kappa/\omega_m$ and for not too large values of the coupling.

Our approach is inspired by Ref.~\cite{CMJ2008}, where the authors consider an approximate description of a BAE optomechanical setup and, besides showing that continuous monitoring can conditionally generate squeezing, they also discuss the implementation of an optimal state-based feedback strategy. Our results complete and considerably extend the analysis reported there. 
Contrary to most treatments of optomechanical feedback protocols~\cite{Doherty99,CMJ2008,Szorkovszky11,Doherty2012PNAS}, our approach does not rely on an effective adiabatic description of the mechanical motion and is not limited to the weak-coupling regime: it enables measurement-based control of quantum squeezing in the good cavity limit where the optical linewidth resolves the sidebands, and in regimes where counter-rotating terms in the optomechanical interaction play a non-negligible role.
Our analysis shows that, even when various limitations are accounted for, feedback control of BAE measurements still provides an effective and versatile tool for deterministic quantum control of mechanical squeezed, and stands out as a useful and promising alternative to reservoir-engineering protocols based on unbalanced two-tone driving~\cite{Kronwald2013}, that have been recently experimentally demonstrated in \cite{Wollman2015,Pirkkalainen2015}.

The present work is structured as follows: in Sec.~\ref{s:Gauss} we introduce the Gaussian framework for describing continuously measured quantum oscillators and linear feedback. In Sec.~\ref{s:BAE} we describe two-tone optomechanical BAE measurements. In Sec.~\ref{s:MarkovFeedback} we assess the performance of Markovian feedback. In particular, we first tackle the time-independent problem (RWA) in Sec.~\ref{s:MarkovRWA} and then include the effects of counter-rotating terms in Sec.~\ref{s:MarkovBeyondRWA}. In Sec.~\ref{s:BayesFeedback} we carry a similar analysis for state-based Bayesian feedback. Finally, in Sec.~\ref{s:Conclusions} contains some final remarks and outlooks.

\section{Conditional evolution and feedback of continuously measured  Gaussian systems}\label{s:Gauss}
We start by reviewing the general formalism describing bosonic Gaussian systems subject to continuous monitoring and modelling the action of a linear feedback, which will be later applied to the case of a two-tone optomechanical BAE measurement.
We consider a system of $n$ quantum harmonic oscillators described by a vector of operators $\hat{\bf r} = (\hat{q}_1, \hat{p}_1, \dots, \hat{q}_n, \hat{p}_n)^{\sf T}$, satisfying the canonical commutation relations
\begin{align}
[\hat{\bf r} ,\hat{\bf r}^{\sf T} ] = i \Omega \,,
\end{align}
where $\Omega = i \bigoplus_{j=1}^n \sigma_y$ is the symplectic form ($\sigma_y$ is the $y$-Pauli matrix). We restrict ourselves to the physical scenario where the oscillators interact via a quadratic Hamiltonian of the form $\hat{\mathcal H}_s = \hat{\bf r}^{\sf T} H \hat{\bf r}/2$, while each of them is linearly coupled to a different Markovian environment. Under these assumptions one can prove that the quantum state $\varrho$ of the oscillators is fully described by the first moments vector $\bar{\bf r} = \Tr[\varrho \hat{\bf r} ]$ and the covariance matrix $\sigmaCM=\Tr[\varrho\{ \hat{\bf r} - \bar{\bf r}, (\hat{\bf r} - \bar{\bf r})^{\sf T} \}]$~\cite{Marco16,AlessioBook}, which evolve according to the equations
\begin{align}
\frac{d\bar{\bf r}}{dt} &= A \bar{\bf r} \,, \label{eq:runc} \\
\frac{d \sigmaCM}{dt} &= A \sigmaCM + \sigmaCM A^{\sf T} + D \,. \label{eq:sigmaunc}
\end{align}
In the equations above, $A$, which known as the drift matrix, depends on $\hat{\mathcal H}_s$ and on the system-environment  interaction, while $D$ denotes the diffusion matrix, which depends on both the system-environment  interaction and the properties of the environment itself, e.g. its temperature (for more details on how to derive these matrices from the actual open-system dynamics we refer to Refs.~\cite{AlessioBook,WisemanDoherty,Marco16}). 

We then assume that the environment is continuously monitored via general-dyne detection \cite{Genoni2014}, yielding a continuous photocurrent
\be
d{\bf y}_t = - \sqrt{2} B^{\sf T} \bar{\bf r}_c \,dt + d{\bf w} \,,
\label{eq:photocurrent}
\ee
where $d{\bf w}$ is a vector of Wiener increments, satisfying $dw_j dw_k = \delta_{jk} dt$, or in a more compact form $\{ d{\bf w}, d{\bf w}^{\sf T} \}/2=\mathbbm{1}\, dt$.  The evolution of the corresponding conditional quantum state $\varrho_c$ is fully described by a stochastic evolution for its first moments, and a deterministic evolution for its covariance matrix, which are respectively given by
\begin{align}
d\bar{\bf r}_c &= A \bar{\bf r}_c\,dt + (E - \sigmaCM_c B) \frac{d{\bf w}}{\sqrt{2}} \,, \label{eq:rc}  \\
\frac{d \sigmaCM_c}{dt} &= A \sigmaCM_c + \sigmaCM_c A^{\sf T} + D - (E - \sigmaCM_c B)(E - \sigmaCM_c B)^{\sf T} \,, \label{eq:sigmac}
\end{align}
where the matrices $E$ and $B$ depends on the specific kind of measurement performed (see Refs.~\cite{AlessioBook,WisemanDoherty,Marco16} for more details). By averaging over all the possible trajectories, that is over all the possible results of the photocurrent $d{\bf y}_t$, one obtains the unconditional state $\varrho_{\sf unc} = \EE[\varrho_c ]$. The averaging procedure leads to increased fluctuations; one can in fact show  that the unconditional covariance matrix takes the form
\be
\sigmaCM_{\sf unc} = \sigmaCM_c + {\boldsymbol \Sigma} \,,
\ee
where  we set $\sigmaCM_{\sf unc}\equiv\sigmaCM$ for clarity and we introduced the excess noise matrix 
\be
\boldsymbol{\Sigma} =  \EE[ \{\bar{\bf r}_c ,\bar{\bf r}_c^{\sf T} \} ] -  \{ \EE[\bar{\bf r}_c] , \EE[\bar{\bf r}_c^{\sf T} ] \}.
\label{eq:excessnoise}
\ee 
Since averaging over all possible measurement outcomes on an ancillary system is equivalent to tracing out the ancillary system, it can be easily checked that $\bar{\bf r}_{\sf unc} = \Tr[\varrho_{\sf unc} \hat{\bf r} ]$ and $\sigmaCM_{\sf unc}$ evolve according to Eqs. (\ref{eq:runc}) and (\ref{eq:sigmaunc}), respectively.

In this work we analyze quantum feedback strategies that are subject to constraints, which reflect some experimental limitations. The goal of feedback is to exploit the information coming from the measurements in order to modify (and optimize) the properties of the unconditional state~\cite{QuantumMeasBook}. 
The feedback is implemented via a Hamiltonian of the form
\begin{align}
\hat{\mathcal{H}}_{\sf fb} = - \hat{\bf r}^{\sf T} \Omega F {\bf u}(t) \,,
\label{eq:feedbackHam}
\end{align}
which corresponds to displacements in the phase space, where the feedback matrix $F$ contains the information on the displacements directions that are allowed, and where the time-dependent feedback signal ${\bf u}(t)$  is chosen according to the feedback strategy. Since the  stochastic term is confined to the first moments [see Eq.~\eqref{eq:rc}], the displacement generated by~\eqref{eq:feedbackHam} is the most general feedback operation that can be implemented. The linear feedback therefore does not affect the conditional evolution of the covariance matrix Eq. (\ref{eq:sigmac}), while the evolution for the first moment vector becomes
\begin{equation}
d\bar{\bf r}_c = A \bar{\bf r}_c\,dt + (E - \sigmaCM_c B) \frac{d{\bf w}}{\sqrt{2}} + F {\bf u}(t) \,dt \,. \label{eq:rcFB}  
\end{equation}
The excess noise matrix ${\boldsymbol \Sigma}_{\sf fb}$ has to be minimized through a suitable choice of the displacements. We stress that the relation $0 \leq {\boldsymbol \Sigma}_{\sf fb} \leq {\boldsymbol \Sigma}$ holds, leading to an unconditional covariance matrix $\sigmaCM_{\sf fb} = \sigmaCM_c + {\boldsymbol \Sigma}_{\sf fb}$. The best result will always correspond to obtain a null matrix ${\boldsymbol \Sigma}_{\sf fb}$, that is to prepare an unconditional state having the same covariance matrix of the conditional one, $\sigmaCM_{\sf fb}=\sigmaCM_c$. 
\section{Optomechanical backaction-evading measurements}\label{s:BAE}
\begin{figure*}[t!]
\includegraphics[width=1\linewidth]{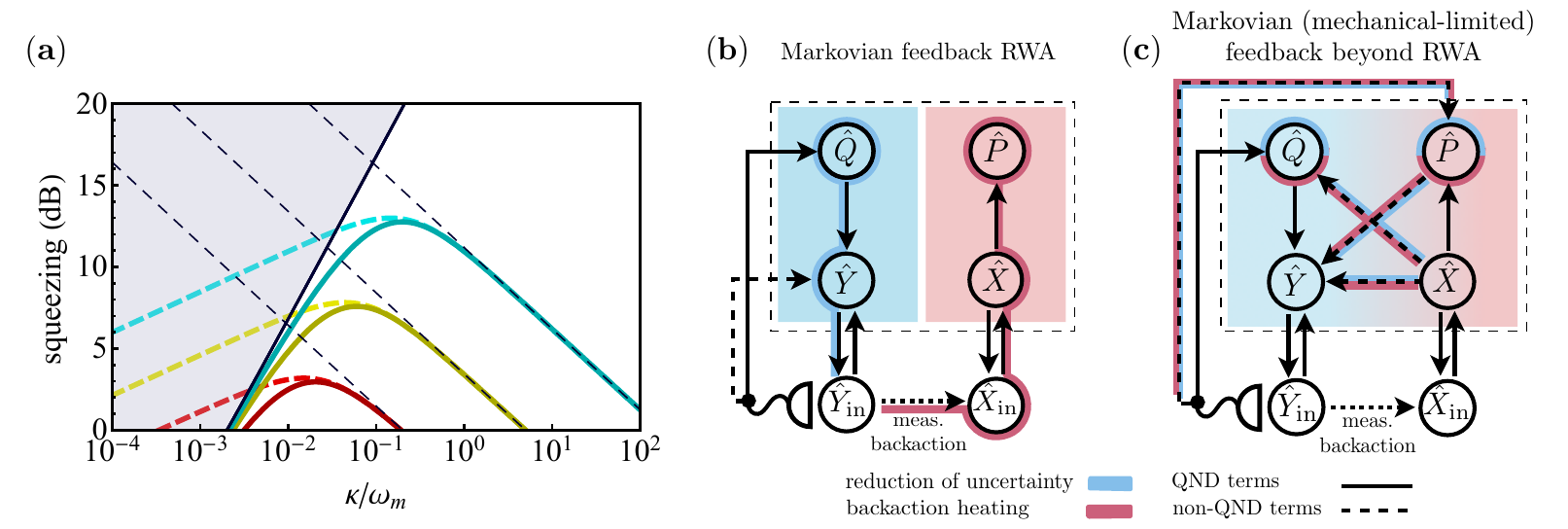}
\caption{{\small ({\bf a}) Squeezing of the mechanical quadrature fluctuations $\langle \Delta\hat{Q}^2 \rangle$ (in decibel) as a function of the sideband parameter $\kappa/\omega_{m}$, for different value of the coupling $g$: $g=0.01\omega_{m}$ (red line), $g=0.05\omega_{m}$ (yellow line), $g=0.3\omega_{m}$ (cyan line). The other values for the parameters are: $\gamma=10^{-4}\omega_{m}$, $\eta=1$, $\bar{n}=10$. Solid lines represent the fluctuations $\langle \Delta \hat{Q}^2\rangle_{\sf fb}$ of the unconditional state obtained via Markovian feedback through the Hamiltonian $\hat{\mathcal{H}}_{\sf fb} = \xi_{m}\hat{P} \,I_{Y}$. Dashed curves represent the fluctuations $\langle \Delta \hat{Q}^2\rangle_c$ of the conditional states [or equivalently of the unconditional state obtained via the optimal feedback Hamiltonian in Eq. (\ref{Hfbottimo})]. The shaded area marks the excluded region beyond the threshold value Eq.~\eqref{SqThr}. Dashed black lines are the predictions of the adiabatic theory Eq.~\eqref{Adiabatic}. ({\bf b}) Effects of measurement-plus-feedback for an ideal BAE measurement. Black arrows describe how quadratures influence each other. By measuring the output phase quadrature, information is extracted from $\hat Q$  (outgoing arrow), while no information can be extracted from $\hat P$ (incoming arrow) where all the backaction goes. Ideal Markovian feedback drives both $\hat Q$ and $\hat Y$. ({\bf c}) Markovian feedback restricted to the mechanical mode (mechanical-limited feedback) and including non-QND terms; these open new paths (dashed arrows) where both backaction and conditioning can spread. The mechanical mode is also subject to (unmonitored) noise from a thermal environment ($\hat{Q}_\mathrm{in}$ and $\hat{P}_\mathrm{in}$), which for convenience is not shown \label{f:FigJoint}}}
\end{figure*}

We consider an optomechanical system composed of a cavity and a mechanical oscillator, respectively described  by bosonic operators $\hat{a}_0$ and $\hat{b}_0$, and with frequencies $\omega_f$ and $\omega_m$. The two oscillators are radiation-pressure coupled with a single-photon coupling $g_0$; the cavity is affected by photon loss with rate $\kappa$, while the mechanical mode interacts with a Markovian phononic bath with decay rate $\gamma$ and a number of thermal phonons $\bar{n}$. We then assume that the cavity is laser-driven at the two frequencies $\omega_\pm = \omega_f \pm \omega_m$ with the same amplitude. The Hamiltonian that describes the system is given by ($\hbar=1$)
\begin{equation}
\label{optom2tonenl}
\hat{\mathcal{H}}_{\sf om}=\hat{\mathcal{H}}_{0}-g_{\tiny 0}\hat{a}_0^{\dagger}\hat{a}_0\left( \hat{b}_0^{\dagger}+\hat{b}_0\right)+\varepsilon(t)\hat{a}_0^{\dagger}+\varepsilon^{\ast}(t)\hat{a}_0,
\end{equation}
where $\hat{\mathcal{H}}_{0}=\omega_f\hat{a}_0^{\dagger}\hat{a}_0+\omega_{m}\hat{b}_0^{\dagger}\hat{b}_0$ is the free Hamiltonian and $\varepsilon(t)=2|\varepsilon|\cos(\omega_{m}t) e^{-i\omega_ft}$ is the driving field. 
Moving to an interaction picture with respect to $\hat{\mathcal{H}}_{0}$ 
and performing a standard linearization procedure~\cite{OptoRev,BowenMilburnBook}, we obtain the interaction Hamiltonian
\begin{equation}\label{optom2tone}
\hat{\mathcal H}_{\sf int}(t)=-g\hat{X}\left[  \hat{Q}\left( 1+\cos(2\omega_{m}t)\right) +\hat{P}\sin(2\omega_{m}t)\right],
\end{equation}
where we have introduced the dimensionless quadratures in the rotating frame $\hat{X}=(\hat{a}+\hat{a}^{\dagger})/\sqrt{2}$, $\hat{Y}=\imath(\hat{a}^{\dagger}-\hat{a})/\sqrt{2}$, for the cavity degree of freedom, and $\hat{Q}=(\hat{b}+\hat{b}^{\dagger})/\sqrt{2}$, $\hat{P}=\imath(\hat{b}^{\dagger}-\hat{b})/\sqrt{2}$ for the mechanical one ($\hat{a}$ and $\hat{b}$ are the annihilation operators in the rotating frame). The parameter $g= g_0 |\varepsilon| \sqrt{\omega_m^2 + \kappa^2/4}$ is now the cavity-enhanced (linearized) coupling strength. This Hamiltonian is composed by a time independent part and an oscillating part. If both conditions (i) $\omega_{m}\gg\kappa$ (\emph{good cavity limit}) and (ii) $\omega_{m}\gg g$ ({\em weak couping}) are fulfilled, the fast oscillating terms in the Hamiltonian quickly average to zero and the Hamiltonian can be written as 
\begin{align}\label{optom2toneRWA}
\hat{\mathcal H}_{\sf int}\simeq -g\hat{X}\hat{Q}.
\end{align}
This Hamiltonian has a quantum non-demolition (QND) form and $\hat{Q}$ is a constant of motion, which makes it is a good QND observable~\cite{Braginsky1,QNDThorne78}. If $\hat{Y}$ is continuously measured, the interaction~\eqref{optom2toneRWA} shunts all the back-action to $\hat{P}$, which is dynamically decoupled from $\hat{Q}$; this mechanism allows to increase the precision of the observable $\hat{Q}$ over time [cf. Fig.~\ref{f:FigJoint} ({\bf b})]. These measurements are called \emph{back-action evading} (BAE) measurements, which are an instance of QND measurement~\cite{Braginsky1,QNDThorne78}. Note that in the following we will use the terms BAE and QND interchangeably. In principle, increasing the system-probe coupling, fluctuations of $\hat{Q}$ can be reduced indefinitely; once passed the SQL, BAE measurements generate a squeezed state for the mechanical oscillator. This ideal scenario is however limited by the presence of unmonitored noise, e.g. from a thermal bath.  

We thus assume that the $\hat{Y}$ quadrature of the cavity is continuously monitored by homodyning the output of the cavity field. Under these assumptions, one can exploit the Gaussian formalism by considering the operator vector $\hat{\bf r} = (\hat{X}, \hat{Y}, \hat{Q}, \hat{P})^{\sf T}$, and the conditional evolution of the quantum state can then be described by Eqs. (\ref{eq:rc}) and (\ref{eq:sigmac}) (see Appendix \ref{a:optomatrices} for more details on the matrices $A$, $D$, $E$, $B$, corresponding to this particular scenario). The analytical solution for the covariance matrix of the stationary conditional state $\sigmaCM_c^{\sf (ss)}$
was derived in Ref.~\cite{MeBAE19}, yielding a variance of the mechanical quadrature $\hat{Q}$
\begin{equation}
\label{RCVbrunelli}
\langle \Delta \hat{Q}^2 \rangle_c=\dfrac{\sqrt{\gamma^{2}+\kappa^{2}+2\zeta}}{16\, g^{2}\eta\kappa}(\zeta+\gamma^{2}-\gamma\sqrt{\gamma^{2}+\kappa^{2}+2\zeta}),
\end{equation}
where $\zeta=\sqrt{\gamma\kappa[16g^{2}\eta(1+2\bar{n})+\gamma\kappa]}$, and $0\leq \eta\leq 1$ is the quantum efficiency of the measurement. In particular it was shown that squeezing, i.e., fluctuations below the vacuum noise $\langle \Delta \hat{Q}^2 \rangle_c < 1/2$, can be in principle generated for a large set of values of the cavity decay rate $\kappa$. 

\section{Markovian feedback}\label{s:MarkovFeedback}
We start our analysis by considering Markovian feedback~\cite{WisemanMilburn94,Wiseman94,Wiseman94Erratum1,Wiseman94Erratum2}. In a Markovian feedback strategy the measured signal is directly fed back to the system. We assume that the feedback signal ${\bf u}(t)$ at time $t$, appearing in the feedback Hamiltonian (\ref{eq:feedbackHam}) depends only on the last photocurrent output ${\bf I}(t) = d{\bf y}_t / dt$, which corresponds to a vanishing delay time in the feedback loop. 
Moreover, we take ${\bf u}(t) = M {\bf I}(t)$, where the matrix $M$ encodes the particular feedback (Markovian) strategy, i.e., it determines how the measured outputs are mixed and weighted when being fed back. The first moment vector evolution~\eqref{eq:rcFB} is then modified as
\be
d\bar{\bf r}_c = \tilde{A}_{\sf m}\bar{\bf r}_c dt + Z \frac{d{\bf w}}{\sqrt{2}} \,, \label{eq:rcFBMarkov}  
\ee
with $\tilde{A}_{\sf m}=(A  - \sqrt{2} F M B^{\sf T})$ and $Z = (E - \sigmaCM_c B) + \sqrt{2} FM$. Notice that the feedback modifies both the drift matrix and the stochastic component; this observation will be especially relevant when compared with the Bayesian strategy in Sec.~\ref{s:BayesFeedback}. By using Ito calculus, we find the following evolution equation for the excess noise matrix (see Appendix~\ref{a:AppendixB} for the derivation)
\begin{align}
\frac{d {\bf \Sigma}_{\sf fb}}{dt} &= \tilde{A}_{\sf m} {\bf \Sigma}_{\sf fb} + {\bf \Sigma}_{\sf fb}\tilde{A}_{\sf m}^{\sf T} + Z Z^{\sf T}  \,. \label{eq:SigmaMarkov}
\end{align}
If one assumes that the feedback matrix $F$ is invertible, namely that displacements are allowed in \emph{all} directions in phase space, one can exploit the residual freedom in the choice of $M$ to completely cancel the stochastic contribution; this situation will be henceforth referred to as the \emph{ideal} case. By doing so one obtains the optimal matrix 
\begin{equation}\label{eq:MarkovianOpt}
M_{\sf opt} = - \frac{F^{-1}(E - \sigmaCM_c^{\sf (ss)} B)}{\sqrt{2}} \,.
\end{equation}
Notice that in the above equation we explicitly opted for canceling the stochastic terms at steady state, since our goal is to maximize the amount of (unconditional) stationary squeezing. One may also make a different choice, e.g. by imposing the stochastic terms to vanish at all times, but this of course would lead to a more onerous kind of feedback.
By enforcing~\eqref{eq:MarkovianOpt}
the excess noise matrix  ${\boldsymbol \Sigma}_{\sf fb}$ will go to zero at steady state, yielding an unconditional state having a covariance matrix equal to the conditional one, i.e., $\sigmaCM_{\sf fb}=\sigmaCM_c^{\sf (ss)}$ (we will always assume that the feedback drift matrix $\tilde{A}_{\sf m}$ is Hurwitz). 

On the other hand, whenever $F$ is not invertible, for example in scenarios where some directions of feedback are not allowed, the stochastic term cannot be identically cancelled; this in turn results in some excess noise. Upon averaging, one indeed obtains a non-zero steady-state excess noise matrix $\boldsymbol \Sigma_{\sf fb}^{\sf (ss)}$, asymptotic solution of the Lyapunov equation (\ref{eq:SigmaMarkov}). This situation will be referred as the \emph{limited} case. Loosely speaking, with Markovian feedback one prioritizes canceling (minimizing) the noise at \emph{steady state}, allowing for some modification in the relaxation dynamics of the system $(\tilde{A}_{\sf m})$. 

\subsection{Mechanical squeezing via Markovian feedback within the RWA approximation}\label{s:MarkovRWA}
We now apply the paradigm just described to the optomechanical setup of Sec.~\ref{s:BAE}. In particular, we start our analysis by  focusing on the scenario where one can apply the RWA  and thus the interaction Hamiltonian is given by Eq. (\ref{optom2toneRWA}). We first address the ideal case,  i.e., we assume a feedback matrix $F= \mathbbm{1}_4$. Under this condition we can derive the optimal Markovian feedback matrix $M_{\sf opt}$ via Eq. (\ref{eq:MarkovianOpt}) and exploit the analytical solution for $\sigmaCM_c^{\sf (ss)}$ available for a two-tone BAE measurement within RWA~\cite{MeBAE19}. The resultaing optimal Markovian feedback Hamiltonian reads
\begin{align}
\hat{\mathcal{H}}_{\sf fb} = \left( \xi_{m}\hat{P}+\xi_{f}\hat{X}\right)I_{Y}(t), 
\label{Hfbottimo}
\end{align} 
where $I_{Y}(t) = -\sqrt{2 \eta \kappa}\, \Tr[\varrho_c \hat{Y}]+ dw/dt$ is the only non-zero element of the photocurrent vector ${\bf I}(t) = d{\bf y}_t / dt$, proportional to the conditional average value $\langle \hat{Y} \rangle_c = \Tr[\varrho_c \hat{Y}]$, and each feedback term is weighted by the factors
\begin{align}
\label{xim}
\xi_{m}&=\dfrac{1}{4g\sqrt{2\kappa\eta}}\left[\gamma^{2}+\zeta-\gamma\sqrt{\kappa^{2}+\gamma^{2}+2\zeta}\right] \,, \\
\label{xic}
\xi_{f}&=\dfrac{1}{2\sqrt{2\kappa\eta}}\left[\kappa+\gamma-\sqrt{\kappa^{2}+\gamma^{2}+2\zeta}\right].
\end{align}
The feedback Hamiltonian~\eqref{Hfbottimo} stabilizes at steady state the \emph{full} optomechanical covariance matrix $\sigmaCM_c^{\sf (ss)}$.
The Hamiltonian consists of a displacement by an amount $\xi_m I_{Y}\,dt$ along the quadrature $\hat{Q}$  and a displacement by $\xi_f I_{Y}\,dt$ along $\hat{Y}$.
The first conclusion to be drawn from Eq.~\eqref{Hfbottimo} is that, even in the ideal constraint-free case, rendering $\sigmaCM_c^{\sf (ss)}$ unconditional requires feedback on both the optical and the mechanical degree of freedom.  A simple explanation for this fact can be given once the effects of both the QND evolution and the measurement are taken into account, as we briefly do in the following. 
In Fig.~\ref{f:FigJoint} ({\bf b}) we represent the Heisenberg evolution of the quadratures as obtained from Eq.~\eqref{optom2toneRWA}, where an arrow connecting two terms means that the variable at the starting point drives the evolution of that at the ending point; the QND interaction entails that $\hat Q$ and $\hat P$ are decoupled.
Following the interaction, the output phase quadrature is measured, which has two main consequences: (i) through the optmechanical coupling, information is acquired about the mechanical quadrature $\hat Q$. This, provided that the values of the photo-current are recorded, \emph{reduces the uncertainty} along $\hat Q$, thus leading to reduced fluctuations and, eventually, to squeezing.  
At the same time, (ii) the measurement \emph{introduces disturbance} (measurement backaction), which directly affects the conjugate quadrature ($\hat X_\mathrm{in}$) and then, through the dynamics, reaches the $\hat P$ quadrature and leads to increased fluctuations (so-called backaction heating). In an ideal BAE measurement the acquisition of information (on $\hat Q$) and the introduction of noise (on $\hat P$) fully decouple. 

Armed with this interpretation, it is now easy to account for the terms featuring in Eq.~\eqref{Hfbottimo}. 
Continuously monitoring the $\hat{Y}_\mathrm{out}$ quadrature causes a stochastic (measurement dependent) displacement along \emph{both} $\hat{Y}$ and $\hat{Q}$, which necessarily accompanies the acquisition of information. Markovian feedback simply removes this effect by counter-displacing both quadratures via the unitary generated by Hamiltonian~\eqref{Hfbottimo}, while displacement of the other two variables ($\hat P$ and $\hat X$) can be disregarded since they are completely decoupled and all the backaction is dumped there. In this way we obtain an unconditional feedback state having the same (optimal) covariance matrix as the conditional one $\sigmaCM_c^{\sf (ss)}$, thus yielding large values of squeezing for the quadrature $\hat{Q}$; some instances  are shown by the dashed lines in Fig.~\ref{f:FigJoint} ({\bf a}). 

We then move to address the limited feedback scenario.
One may naively think that, since the figure of merit we consider pertains only a part of $\sigmaCM_c^{\sf (ss)}$, a \emph{single} displacement would suffice to render unconditional the sub-block we are interested in. 
However, from Fig.~\ref{f:FigJoint} ({\bf b}), we see that the measurement correlates $\hat Q$ and $\hat Y$,
inasmuch as it jointly reduces their uncertainty. Therefore, local operations on one mode will in general affect local properties on the other (such as squeezing).
Moreover, the QND coupling imposes a fundamental asymmetry between the two quadratures, whereby noise from $\hat Q$ drives $\hat Y$ but not the other way round.
Therefore, we can already conclude that by limiting the feedback to a single displacement in general we will  not recover the optimal squeezing. In particular, {\em cavity-limited} feedback, i.e., implementing only the optical part of $\hat{\mathcal{H}}_{\sf fb}$, yields poor results, since the feedback acts `downstream' with respect to the QND evolution [any stochastic term driving $\hat Q$ will also drive $\hat Y$, cf. Fig.~\ref{f:FigJoint} ({\bf b})]. This intuition can be made rigorous by neglecting the mechanical term in Eq. (\ref{Hfbottimo}), thus obtaining $\hat{\mathcal{H}}_{\sf fb, f} = \xi_{f}\hat{X} \,I_{Y}(t)$.
The corresponding steady-state excess noise matrix $\boldsymbol \Sigma_{\sf fb}$ can be obtained via Eq. (\ref{eq:SigmaMarkov}) by considering the feedback matrix $F_{\sf f} = \mathbbm{1}_2 \oplus \mathbb{0}_2$ (with $\mathbb{0}_d$ denoting a square matrix of dimension $d$ with all elements equal to zero). Under this restriction the quadrature fluctuations of the unconditional (feedback) state $\langle \Delta \hat{Q}^2 \rangle_{\sf fb}$ are only slightly reduced below the case with no feedback, and consequently no squeezing can be observed. 
 
On the other hand, if we consider {\em mechanical-limited} feedback via the Hamiltonian $\hat{\mathcal{H}}_{\sf fb, m} = \xi_{m}\hat{P} \,I_{Y}(t)$, we find much better results. Again, this result can be expected as the feedback now acts  `upstream'  with respect to the QND evolution [cf. Fig.~\ref{f:FigJoint} ({\bf b})]. The steady-state excess noise matrix can be obtained by considering a  feedback matrix $F_{\sf m} = \mathbb{0}_2 \oplus \mathbbm{1}_2$ and the corresponding values of $\langle \Delta \hat{Q}^2 \rangle_{\sf fb}$ are plotted in Fig.~\ref{f:FigJoint} ({\bf a})  [expressed in $-10\log_{10}\langle \Delta \hat{Q}^2 \rangle$ Decibel (dB)], alongside the optimal values of the conditional state $\langle \Delta \hat{Q}^2 \rangle_{\sf c}$.  Since the stochastic contribution from the cavity field is not removed, averaging determines increased fluctuations, i.e., $\langle \Delta \hat{Q}^2 \rangle_{\sf fb}\ge\langle \Delta \hat{Q}^2 \rangle_{\sf c}$. Physically, the reason why mechanical-limited feedback remains suboptimal is that, although the feedback removes the stochastic term for the evolution of the $\hat{Q}$ quadrature,
the displacement of $\hat{Q}$ is proportional to the photocurrent, and thus to $\langle\hat{Y}\rangle_c$; as the quadratures $\hat{Q}$ and $\hat{Y}$ are correlated, the fluctuations of $\hat{Y}$ (that have not been reduced by the feedback) will have a non-zero effect on the fluctuations of $\hat{Q}$, reducing the amount of squeezing that one can generate.

In Fig.~\ref{f:FigJoint} ({\bf a}) we can clearly observe two distinct regimes for stationary squeezing: in the {\em bad-cavity} limit ($\kappa \gg \omega_m$) mechanical-limited feedback turns out to be optimal, while in the {\em good-cavity} limit ($\kappa \ll \omega_m$) we obtain worse results. Although an analytic expression of $\langle \Delta \hat{Q}^2 \rangle_{\sf fb}$ is available, it is too cumbersome to be reported here. We instead now derive simple expressions for these two limits.

 In the good cavity limit mechanical-limited feedback leads to a \emph{universal} upper bound on the amount of squeezing attainable; by universal we mean that the value is independent of both the strength of the coupling and the detection efficiency. In this limit, the stochastic displacement along $\hat Y$ affects long-lived cavity photons, so averaging upon it leads to comparatively larger excess noise. Expressing $\langle \Delta \hat{Q}^2 \rangle_{\sf fb}$ in terms of the multi-photon cooperativity $\mathcal{C}=4 g^2/\kappa \gamma$ and keeping  the leading term in the expansion $\mathcal{C}\gg1$, we obtain the threshold value
\begin{equation}\label{SqThr}
\langle \Delta \hat{Q}^2 \rangle_{\sf thr}=\frac{\gamma (2 \bar n+1)}{\gamma+\kappa}\,,
\end{equation}
which corresponds to the black line in Fig.~\ref{f:FigJoint} ({\bf a}).
In particular, from Eq.~\eqref{SqThr} it follows that there exists an excluded region of sideband values $\kappa/\omega_m< 2\bar n /\mathcal{Q}_m$, (with $\mathcal{Q}_m=\omega_m/\gamma$ being the mechanical quality factor) where squeezing cannot be attained for any value of the coupling strength; this is indicated by the shaded region in figure. This represents a nontrivial prediction of our framework, as it sets a fundamental lower bound on the achievable precision via Markovian feedback. 

In the bad-cavity limit we see that mechanical-limited feedback achieves optimal squeezing  and the inequality $\langle \Delta \hat{Q}^2 \rangle_{\sf fb}\ge\langle \Delta \hat{Q}^2 \rangle_{\sf c}$ is saturated. This behaviour can be simply understood by realizing that for a large enough linewidth, the photon lifetime inside the cavity is so short that optical feedback becomes inconsequential. In this limit the cavity field can be adiabatically eliminated, one obtains  $\langle \Delta \hat{Q}^2 \rangle_{\sf fb}=\langle \Delta \hat{Q}^2 \rangle_{\sf c}\approx \langle \Delta \hat{Q}^2 \rangle_{\sf ad}$, where the last quantity has the following expression
\begin{equation}\label{Adiabatic}
\langle \Delta \hat{Q}^2 \rangle_{\sf ad}=\frac{\sqrt{1+4 \eta \mathcal{C} (2 \bar n+1)}-1}{4 \mathcal{C} \eta }\,,
\end{equation}
and corresponds to the black dashed lines in the plot (for further details about this behaviour see~\cite{MeBAE19}). The adiabatic prediction dramatically fails moving towards good cavity limit, which is where most experiments take place. This expresses the inadequacy of adiabatic treatment of measurement-based squeezing available so far.   

Let us now make a crucial observation regarding the mechanical feedback term $\hat{\mathcal{H}}_{\sf fb,m} = \xi_{m}\hat{P} \,I_{Y}(t)$. We remind that we are working in interaction picture with respect to the free Hamiltonian. If we go back to the laboratory frame operators $\hat{Q}_0$ and $\hat{P}_0$, corresponding to the actual position and momentum of the mechanical oscillator, we obtain the feedback Hamiltonian
\begin{align}
\hat{\mathcal{H}}_{\sf fb,m} &= \xi_{m}\left( \cos(\omega_{m}t)\hat{P}_0-\sin(\omega_{m}t)\hat{Q}_0\right)I_{Y}(t).
\end{align}
The term proportional to the position operator $\hat{Q}_0$ corresponds to a mechanical force. In clamped resonators, this can be implemented via piezoelectric actuators~\cite{Poggio07} or via radiation-pressure force from an auxiliary laser beam (not coupled to the cavity mode)~\cite{Rossi2018}; mechanical feedback forces have been implemented also in levitated charged nanoparticles via electrodes placed in the vicinity of the particle for cooling its centre-of-mass motion~\cite{Tebbenjohanns19,Conangla19,Iwasaki19}. On the other hand, terms proportional to momentum are notoriously more challenging to implement~\cite{Doherty99}. Therefore it is physically motivated to assume that the feedback action is implemented only by means a (possibly time-dependent) force on the mechanical oscillator.  We refer to this scenario as {\em force-limited} feedback. Neglecting the terms proportional to $\hat{P}_0$ in the Hamiltonian above and going back to the rotating quadratures, we obtain
\begin{align}
\label{Hfblima}
\hat{\mathcal H}_{\sf fb,force}^{\sf (1)}&= -\xi_m \sin(\omega_{m}t)\hat{Q}_0 \, I_{Y}(t) \,,\nonumber \\
&=\frac{1}{2}\xi_{m}\left(\hat{P}-\cos(2\omega_{m}t)\hat{P}-\sin(2\omega_{m}t)\hat{Q}\right)I_{Y}(t)\,, \nonumber
\\& = \frac{1}{2}\hat{\mathcal H}_{\sf fb,m}+\hat{\mathcal H}_{\sf rot}.
\end{align}
This phyiscally-constrained feedback Hamiltonian is composed of a time-independent part, equal to half the optimal feedback Hamiltonian in Eq. (\ref{Hfbottimo}), plus a term $\hat{\mathcal H}_{\text{rot}}=-\frac{1}{2}\xi_{m}(\cos(2\omega_{m}t)\hat{P}+\sin(2\omega_{m}t)\hat{Q}) I_Y(t)$, with elements rotating at frequency $2\omega_m$. These terms are depicted in Fig.~\ref{f:FigJoint} ({\bf c}), from which we see that the feedback now drives \emph{both} mechanical quadratures. Of course, we may as well assume to be able to double the feedback signal ${\bf u}(t)$, thus obtaining a force-feedback Hamiltonian 
\begin{align}
\hat{\mathcal H}_{\sf fb,force}^{\sf (2)} &= -2 \xi_m \sin(\omega_{m}t)\hat{Q}_0 \, I_{Y}(t) \nonumber \\
&= \hat{\mathcal H}_{\sf fb,m}+2 \hat{\mathcal H}_{\sf rot}.
\label{eq:MarkovianLimited}
\end{align}
As long as we are working within the RWA, the two choices~\eqref{Hfblima},~\eqref{eq:MarkovianLimited} are equally viable. It would then seem that the results shown in Fig.~\ref{f:FigJoint} ({\bf a}) can always be obtained by simply implementing a time-dependent feedback force.
This seeming contradiction can be cleared by taking a closer inspection at the RWA under Markovian feedback.
In the presence of feedback, besides the weak coupling condition needed to cast the two-tone optomechanical Eq.~\eqref{optom2tone} into a QND form, also the condition $\vert\xi_m I_{Y}(t)\vert\ll\omega_m$ needs to be fulfilled in order for the RWA to be valid. Substituting the expression for the homodyne current, the latter condition splits in two parts; the first part yields $\omega_m\gg\xi_m \sqrt{2 \eta \kappa} \langle\hat{Y}\rangle_c\approx \sqrt{\eta \kappa\gamma(2\bar n+1)} \langle\hat{Y}\rangle_c$, where the last approximation holds for high-Q mechanical resonators, while the second part is $\vert\xi_m dw/dt\vert\ll\omega_m$. However, at any given instant the current is dominated by white noise contribution, which takes unbounded values, so that the second condition cannot be fulfilled. Therefore, strictly speaking the RWA is never fully justified when dealing with Markovian feedback and counter-rotating terms cannot be overlooked. 
\subsection{Mechanical squeezing via limited Markovian feedback beyond the RWA approximation}\label{s:MarkovBeyondRWA}
\begin{figure}
\begin{center}
\includegraphics[scale=0.75]{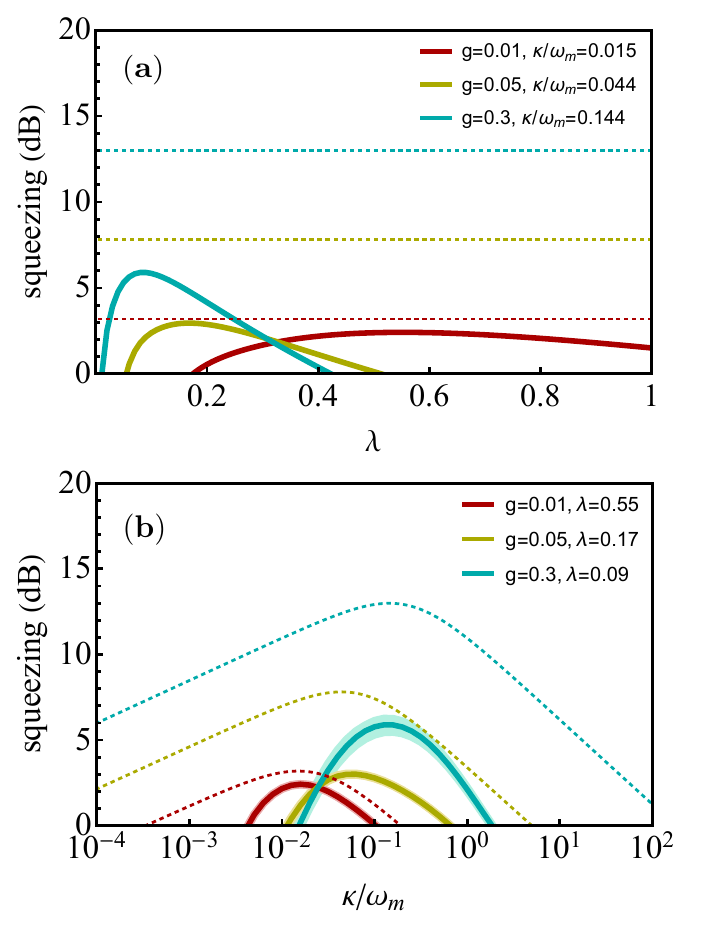}
\caption{{\small {\bf (a)}: squeezing of the mechanical quadrature fluctuations $\langle \Delta\hat{Q}^2 \rangle$  (in decibel) obtained via the non-optimal Markovian feedback matrix in Eq. (\ref{HfblimNoRWA3}), as a function of the 
Hamiltonian parameter $\lambda$ and for three couples of parameters $(g,\kappa)$. The dotted lines denotes the corresponding fluctuations $\langle \Delta\hat{Q}^2 \rangle_{\sf c}$ of the conditional states for the same values of the parameters. Notice that the couples $(g,\kappa)$ has been chosen in order to pick the maximum of squeezing obtainable from the conditional states, at fixed coupling $g$ and varying $\kappa$ as in Fig. \ref{f:FigJoint} ({\bf a}). \\
{\bf (b)}: squeezing of the mechanical quadrature fluctuations $\langle \Delta\hat{Q}^2 \rangle$ (in decibel) obtained via the non-optimal Markovian feedback matrix in Eq. (\ref{HfblimNoRWA3}), as a function of the sideband parameter $\kappa/\omega_{m}$, for different values of the coupling $g$, and by choosing  the values of $\lambda$ maximizing the squeezing as shown in the upper panel. As above, the dotted lines denotes the corresponding fluctuations $\langle \Delta\hat{Q}^2 \rangle_{\sf c}$ of the conditional states.
\\
In both panels the values for the other parameters are fixed to: $\gamma=10^{-4}\omega_{m}$, $\eta=1$, $\bar{n}=10$. }}
\label{f:Markovian_bRWA}
\end{center}
\end{figure}
We now include the effect of counter-rotating terms that are neglected under RWA.  As explained above, this analysis is crucial to assess the performances of Markovian feedback control, in particular when restricting the feedback Hamiltonian to be a force.
Due to mixing between the quadratures, a rotating frame where the equations of motion become time-independent can no longer be found~\cite{Mari2009,DanielBAE}, nor a closed expression for the conditional state.
 In this case one has to consider the time-dependent drift matrix corresponding to the interaction Hamiltonian (\ref{optom2tone}) and numerically integrate the evolution equation for the conditional covariance matrix (\ref{eq:sigmac}). 
However, we can still get a qualitative picture of the effects brought about by the counter-rotating terms by looking at Fig.~\ref{f:FigJoint} ({\bf c}). counter-rotating terms open new paths (dashed arrows) between the quadratures, where both backaction and conditioning can spread. As we can see, measurement backaction is no longer confined to $\hat P$ but now reaches both quadratures, which entails a reduction of the amount of squeezing with respect to the ideal BAE regime. At the same time, the measurement now acquires information about both mechanical quadratures, which entails  that the conditional state gets purified~\cite{MeBAE19}. Finally, information is simultaneously acquired about both the cavity and the mechanics (multiple arrows incoming at $\hat Y$), which implies that stronger correlations between cavity and mechanics are built. Although our analysis will be focused on the reduction of squeezing, the point to be stressed here is that, contrary to common wisdom, counter-rotating terms are not only tied to detrimental effects but can actually be beneficial for conditional state preparation. This offers a further motivation for implementing feedback in BAE measurements beyond the weak coupling regime.

We start by considering the optimal (time-dependent) Markovian matrix $M_{\sf opt}$, namely all phase-space directions are allowed. It can be numerically derived via Eq. (\ref{eq:MarkovianOpt}), by choosing $F=\mathbbm{1}_4$ 
and by replacing $\sigmaCM_c^{\sf (ss)}$ with the numerical solution of Eq. (\ref{eq:sigmac}). The corresponding optimal Markovian feedback Hamiltonian reads
\begin{align}
\label{HfblimNoRWA1}
\hat{\mathcal H}_{\sf fb}= (m_{32}\hat{P} - m_{42}\hat{Q} - m_{22}\hat{X} + m_{12}\hat{Y})I_{Y},
\end{align}
where the elements $m_{jk}$ appearing are the only non-zero elements of $M_{\sf opt}$. Compared to Eq.~\eqref{Hfbottimo}, $\hat{\mathcal H}_{\sf fb}$ now generates displacements also along $\hat P$ and $\hat X$.
As done before, we now purposedly neglect the cavity field terms (mechanical-limited feedback) and move back to the laboratory frame operators
\begin{align}
\label{HfblimNoRWA2}
\hat{\mathcal H}_{\sf fb,m} &= (m_{32}\hat{P} - m_{42}\hat{Q})I_{Y}, \nonumber \\
&= \left[m_{32}\left( \hat{P}_0 \cos(\omega_m t) - \hat{Q}_0 \sin(\omega_m t) \right) +  \right. \nonumber \\
& \,\,\, \left. - m_{42}\left(\hat{Q}_0 \cos(\omega_m t) + \hat{P}_0 \sin(\omega_m t) \right)\right] I_{Y}. 
\end{align}
We further restrict the feedback to act as  a force on the oscillator (force-limited feedback), and following the line of reasoning 
of the previous section we consider the following feedback Hamiltonian
\begin{align}
\label{HfblimNoRWA3}
\hat{\mathcal H}_{\sf fb,force}(\lambda)=& - 2\lambda \, [m_{4,2} \cos(\omega_m t) + m_{3,2} \sin(\omega_m t)] \hat{Q}_0\,  I_{Y}\,, \nonumber\\
&= \lambda \, \hat{\mathcal H}_{\sf fb,m} + 2\lambda \, \hat{\mathcal H}_{\sf rot} \,
\end{align}
where $\hat{\mathcal H}_{\sf fb,m}$ here denotes the mechanical-restricted optimal Hamiltonian in Eq. (\ref{HfblimNoRWA2}) and where $\hat{\mathcal H}_{\sf rot}$ contains elements rotating a twice the mechanical frequency $2 \omega_m$; these additional contributions are sketched in Fig.~\ref{f:FigJoint} ({\bf c}). The free real parameter $\lambda$  allows to interpolate between two cases: choosing $\lambda=1/2$ corresponds to consider only the elements proportional to $\hat{Q}_0$, while for $\lambda=1$ we double the feedback signal, re-obtaining the optimal Hamiltonian (\ref{HfblimNoRWA2}) plus a \emph{larger} counter-rotating term $2\hat{\mathcal H}_{\sf rot}$. In our formalism, this choice corresponds to multiplying the optimal feedback matrix $M_{\sf opt}$ with a  time-dependent limited feedback matrix of the form $F_{m} = (2\lambda) \left(\mathbb{0}_2 \oplus W \right)$, with 
\begin{align}
W &= R F_{\sf force} R^{\sf T} \,, \nonumber \\
&= \left(
\begin{array}{c c}
\sin^2 (\omega_m t) & \sin (\omega_m t) \cos(\omega_m t) \\
 \sin (\omega_m t) \cos(\omega_m t) &  \cos^2(\omega_m t)
\end{array}
\right) \,, \label{eq:matrixW}
\end{align}
where $R$ is the rotation matrix by an angle $\omega_m t$ and $F_{\sf force}= {\rm diag}(0,1)$ is the feedback matrix  corresponding to feedback displacements along $\hat{P}_0$ axis only.

We can now numerically integrate Eq.~\eqref{eq:SigmaMarkov} for the excess noise matrix $\boldsymbol \Sigma_{\sf fb}$, and evaluate the corresponding fluctuation $\langle \Delta \hat{Q}^2 \rangle_{\sf fb}$, averaged over a period (the integration is carried out until $\sigmaCM_c$ and $\boldsymbol \Sigma_{\sf fb}$ reach a time-periodic steady state). In Fig.~\ref{f:Markovian_bRWA} ({\bf a}), we have plotted $\langle \Delta \hat{Q}^2 \rangle_{\sf fb}$ as a function of $\lambda$ for the three choices of the parameters $(g,\kappa)$ yielding the maximum values of squeezing observed in in Fig.~\ref{f:FigJoint} ({\bf a}). As expected, for larger values of $g$ we observe a larger reduction of the squeezing. Remarkably, we also observe that the counter-rotating terms appearing in the feedback Hamiltonian play a major role. Indeed, the optimal value of $\lambda$ maximizing the squeezing in general corresponds to neither $\lambda=1/2$ nor $\lambda=1$. We thus conclude that none of the two approaches discussed at the end of the previous section corresponds to the optimal choice. In particular, as we may now expect, we notice how when increasing the opto-mechanical coupling $g$, one should choose smaller values of $\lambda$: the counter-rotating terms in the feedback Hamiltonian in the {\em strong-coupling} regime have in fact a major role. However it is also important to remark that for relatively small values of $g$, applying this physically constrained feedback strategy yield only a small reduction in the squeezing that one can obtain unconditionally, respect to the one obtained via the continuously-monitored conditional states. 

Finally, in Fig.~\ref{f:Markovian_bRWA} ({\bf b}) we show a comparison between RWA and full Hamiltonian for a particular choice of the experimental parameters values, and by choosing the corresponding optimal value of $\lambda$. The solid curve shows the mean unconditional squeezing (averaged over one mechanical period) and the shaded area extends between the minimum and maximum value of squeezing. We see that Markovian feedback enforces a more conservative condition for RWA, which strictly speaking is never fulfilled. This is confirmed by looking at the weak coupling instances where, even when the RWA on the QND Hamiltonian~\eqref{optom2toneRWA} provides an excellent approximation, $\langle \Delta \hat{Q}^2 \rangle_{\sf fb}$ is still appreciably smaller than the corresponding $\langle \Delta \hat{Q}^2 \rangle_{\sf c}$. 
Our results show how counter-rotating terms, that have been so far neglected, have a non trivial effect even in weak coupling regime, where one would expect RWA to be excellent approximation, reinforcing the need of accounting them in the assessment of these control strategies.

\section{Bayesian feedback}\label{s:BayesFeedback}
We now assume that the feedback signal ${\bf u}(t)$ can be chosen by taking into account the whole measurement results, that is all the values of the photocurrent $d{\bf y}_{s}$, with $0\leq s \leq t$; these are used to estimate properties of the conditional state, which are in turn exploited in the feedback step. This kind of non-Markovian feedback is typically referred to as {\em state-based feedback} or {\em Bayesian feedback}, as determining the conditioned state of the quantum system from classical photo-current corresponds indeed to a quantum version of the classical Bayesian update \cite{WisemanBayesian2002,QuantumMeasBook}.
Here we will focus on the minimization of {\em quadratic} cost function defined as 
\begin{align}
h(t) =\mathbbm{E} [\langle \hat{\bf r}^{\sf T} S \hat{\bf r} \rangle_c + {\bf u}^{\sf T} \Xi {\bf u} ]\,,\label{eq:costfunction}
\end{align}
that one typically integrates over a certain time interval $\bar{h} = \int_0^{\bar t} h(s) ds$. In the following we will be interested in optimizing this cost function at the (possibly time periodic) steady state; we will then consider the infinite-time limit $h_{\sf ss} = \lim_{t\rightarrow\infty} h(t)$. The positive semi-definite matrix $S \geq 0$ sets  the particular property of the system that we want to minimize, while the positive-definite matrix $\Xi>0$ quantifies the cost of the linear driving ${\bf u}(t)$ that we are implementing with our feedback strategy. Under these assumptions we are dealing with the paradigm of linear-quadratic-Gaussian (LQG) control~\cite{QuantumMeasBook}. This is indeed a well-known classical optimal control problem, which is well suited for Gaussian quantum systems. It has been previously applied to optomechanical systems, e.g. to cool the mechanical oscillator~~\cite{Doherty99,Doherty2012PNAS,Hofer15}, or harness the optomechanical entanglement generated in the blue-detuned regime for various state preparation tasks~\cite{Hofer15}. Moreover, a crucial ingredient of LQG control, namely optimal quantum state estimation (corresponding to the classical Kalman filter), has been recently demonstrated for both mechanically compliant resonators~\cite{Wieczorek15,Rossi19} and levitated nanoparticles~\cite{Setter18,Liao19}. According to LQG control, the solution minimizing Eq.~\eqref{eq:costfunction} is obtained by considering a feedback signal depending linearly on the first moment vector
\begin{equation}\label{SignalBayes}
{\bf u}(t) = - K(t) \bar{\bf r}_c \,,
\end{equation}
such that the evolution of the first moments is rewritten as
\be
d\bar{\bf r}_c = \tilde{A}_{\sf b}\bar{\bf r}_c \,dt + L \frac{d{\bf w}}{\sqrt{2}} \,.
\label{eq:rcFBBayes}
\ee
with $\tilde{A}_{\sf b} = (A - F K(t) )$ and $L = (E - \sigmaCM_c B)$. We stress that Bayes feedback~\eqref{SignalBayes} employs only the mean values (sometimes referred to as the `estimates') of the conditional
state, i.e., the feedback signal is noiseless. One further proves that the matrix $K_{\sf opt}$, optimizing the steady-state cost function $h_{\sf ss}$ reads
\begin{equation}
K_{\sf opt} = \Xi^{-1} F^{\sf T} Y \,
\label{eq:Kopt}
\end{equation}
where $Y$ is the solution of the (homogeneous) Riccati equation 
\begin{align}
0 = A^{\sf T} Y + Y A + S  - Y F \Xi^{-1} F^{\sf T} Y  \,.
\label{eq:ypsilon}
\end{align}
In this case the evolution for the feedback excess noise matrix is given by (see Appendix \ref{a:AppendixB} for details on the derivation)
\begin{align}
\frac{d \boldsymbol{\Sigma}_{\sf fb}}{dt} &= \tilde{A}_{\sf b}\boldsymbol{\Sigma}_{\sf fb} + \boldsymbol{\Sigma}_{\sf fb} \tilde{A}_{\sf b}^{\sf T} + LL^{\sf T} \,.
\label{eq:SigmaBayes}
\end{align}
By solving the corresponding Lyapunov equation, one can thus calculate the steady-state excess noise matrix $\boldsymbol{\Sigma}_{\sf fb}^{\sf ss}$ and assess the performance of the feedback strategy.\\

By comparing Eqs. (\ref{eq:SigmaMarkov}) and (\ref{eq:SigmaBayes}), we notice an important difference in the working principles of Markovian and Bayesian feedback. While both strategies change the {\em drift matrix} $A$, adding a damping term to the first moments, and consequently to the excess noise matrix $\Sigma_{\sf fb}$, Markovian strategies also aim to reduce the {\em diffusion term} in the Lyapunov equation (\ref{eq:SigmaMarkov}), cancelling it  in the optimal scenario and yielding an unconditional feedback covariance matrix $\sigmaCM_{\sf fb}=\sigmaCM_{\sf c}$. It is also important to remark that, as mentioned in the previous section, Markovian feedback is more {\em expensive}: the white noise term $d{\bf w}/dt$, entering into the feedback signal ${\bf u}(t)$ via the photocurrent, yields indeed a diverging average $\mathbbm{E}[(F{\bf u}(t))^{\sf T} F{\bf u}(t)]$. On the other hand, Bayesian LQG feedback strategies, by fixing a non-zero cost on the feedback displacement, involve always a finite average signal, and in this sense, as we will see in the following, it is of interest in practical implementations. The average feedback at steady-state can be evaluated via the formula~\cite{QuantumMeasBook}
\begin{equation}
\lim_{t \to \infty} \mathbbm{E}[(F{\bf u}(t))^{\sf T} F {\bf u}(t)] = \Tr[F K_{\sf opt} \boldsymbol{\Sigma}_{\sf fb}^{\sf (ss)} K_{\sf opt}^{\sf T} F^{\sf T}  ]. \label{eq:averageFB}
\end{equation}
This quantity is in general finite, as it diverges only by taking the limit of zero cost matrix $\Xi \to 0$; in this limit one is supposed to implement an infinite damping matrix $K_{\sf opt}$, and, by considering a full-rank feedback matrix $F=\mathbbm{1}$, one obtains a steady-state zero excess noise matrix and thus the optimal result $\sigmaCM_{\sf fb}^{\sf (ss)} = \sigmaCM_c^{\sf(ss)}$. \\

\subsection{Mechanical squeezing via Bayesian feedback within the RWA approximation}
We now apply this formalism to our optomehcanical setup. Our goal is to minimize the steady-state fluctuations of the $\hat{Q}$ quadrature, with a non-zero cost on the feedback displacement. In terms of the figure of merit $h_{\sf ss}$, this is obtained by the choosing $S = {\rm diag}(0,0,1,0)$ and $\Xi = \chi \mathbbm{1}_4$, where the parameter $\chi>0$ weights the cost of the overall feedback with respect to the property to be optimized (squeezing along $\hat{Q}$). By considering the matrix $
\Xi$ proportional to the identity we are assuming equal cost for all phase-space directions of feedback displacement. Nevertheless, as done in the Markovian case, we will indirectly impose infinite cost along some directions in phase-space by choosing non-full rank feedback matrices $F$. \\
\begin{figure*}[t!]
\centering
\includegraphics[width=1\linewidth]{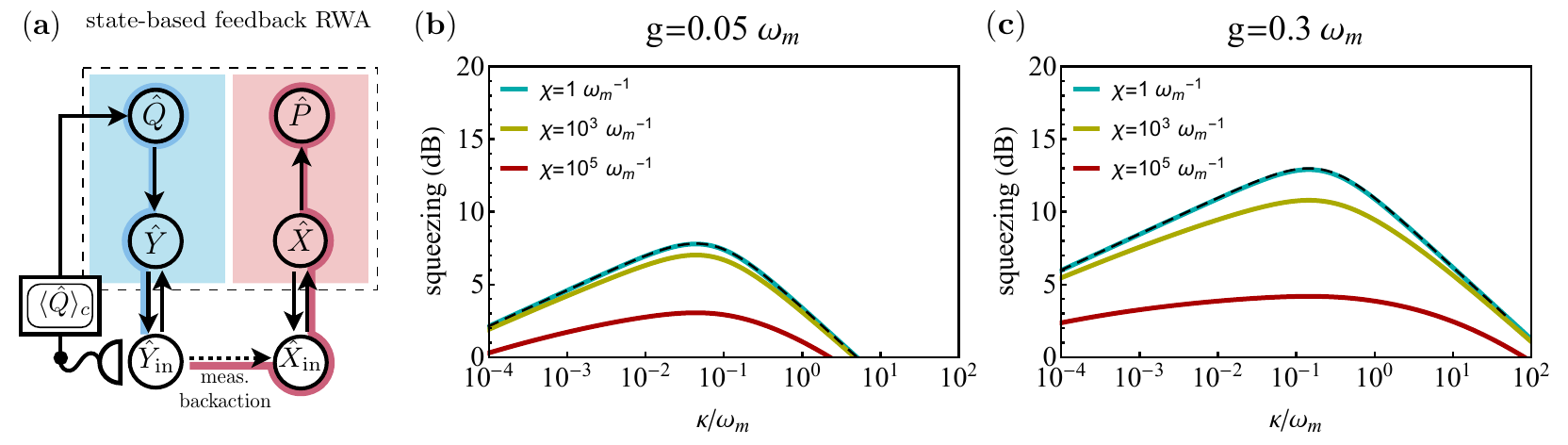}
\caption{\small{({\bf a}) Diagram depicting how quadratures influence each other in the case of ideal state-based (Bayesian) feedback; all conventions are as in Fig.~\ref{f:FigJoint}. Mechanical position variance in terms of squeezing factor (dB) as a function of the sideband parameter $\kappa/\omega_{m}$, for three different values of cost parameter (solid lines). The dashed black line denotes the optimal squeezing obtained via the conditional evolution within RWA approximation, corresponding to Eq. (\ref{RCVbrunelli})).
Different panels correspond to different values of the coupling $g$ [({\bf b}), $g=0.05 \omega_m$; ({\bf c}), $g=0.3 \omega_m$], while the other parameters are fixed as following: $\eta=1$, $\bar{n}=10$, $\gamma=10^{-4}\omega_{m}$.}}
\label{f:OptBayesianFB}
\end{figure*}
We start by considering the ideal feedback matrix $F=\mathbbm{1}_4$. By applying the formulas above, one can analytically obtain the optimal matrix $K_{\sf opt}$ and the corresponding feedback Hamiltonian, which reads
\begin{align}
\hat{\mathcal H}_{\sf fb,b} &=\dfrac{\langle \hat{Q}\rangle_c}{2}\left( \gamma-\sqrt{\dfrac{4}{\chi}+\gamma^{2}}\right) \hat{P}\,, \nonumber \\
&= - \beta \langle \hat{Q}\rangle_c  \hat{P} \, ,
\label{Hfbbay}
\end{align}
where $\beta$ is a positive real parameter, monotonically decreasing with the parameter $\chi$ (the smaller the feedback cost $\chi$, the larger the damping of the conditional average value $ \langle \hat{Q}\rangle_c$). The first thing to be noticed is that the ideal  feedback Hamiltonian contains  no cavity terms, which in this framework turn out to be unnecessary. 
Indeed, as explained before, Bayesian feedback corresponds only to an attenuation of the first moment vector and consequently on the excess noise matrix. As our goal is to reduce the fluctuations of the quadrature mechanical operator $\hat{Q}$, the feedback Hamiltonian does lead to a damping only in this direction of phase-space, and other feedback operations are not necessary. We also notice that, as expected, in the limit of either zero or infinite cost parameter $\chi$, we implement respectively an infinite or zero damping. Finally we observe that the feedback Hamiltonian, apart from the cost parameter $\chi$, only depends on the parameter $\gamma$, since all the salient information is incapsulated in the average value $\langle \hat{Q} \rangle_c$. 

Solving Eq. (\ref{eq:SigmaBayes}), we derive an analytical expression for the excess noise at steady state for the mechanical operator $\hat{Q}$
\begin{equation}
\label{sigbbay}
\Sigma^{ss}_Q=\dfrac{\sqrt{\chi}}{\sqrt{4+\gamma^{2}\chi}}\frac{(\gamma^{2}+\zeta-\gamma\sqrt{\kappa^{2}+\gamma^{2}+2\zeta})^{2}}{16g^{2}\eta\kappa}.
\end{equation}
In Fig. (\ref{f:OptBayesianFB}) we plot the behavior of the mechanical squeezing (in terms of dB) as a function of the sideband parameter $\kappa/\omega_m$. We observe that increasing the cost parameter $\chi$ or decreasing the optomechanical coupling $g$ have a different effect on the steady-state squeezing achievable. In particular, increasing $\chi$ penalizes the ``intermediate'' sideband values for which the squeezing is maximum, while the range for which squeezing can be observed remains almost unchanged compared to the optimal conditional states. 

It is important to stress that Bayesian feedback allows to obtain nearly optimal squeezing also in the {\em good cavity} limit, notwithstanding the fact that the optimal feedback Hamiltonian acts \emph{on the mechanical oscillator only}. This shows a fundamental difference respect to the Markovian scenario, where mechanical-limited feedback yields a large reduction of the squeezing achievable for $\kappa/\omega_m \ll 1$ and to a threshold value under which no squeezing can be observed. Furthermore, the fact that, for a fixed cost $\chi$, a larger coupling constant $g$ implies larger deviations from the optimal conditional squeezing, strongly suggests that larger coupling values demand larger values of the feedback signal ${\bf u}(t)$. \\

\subsection{Mechanical squeezing via Bayesian feedback beyond the RWA approximation}
We now consider the effect of  counter-rotating terms in the optomechanical Hamiltonian (\ref{optom2tone}). Moreover, similarly to the Markovian case, we also focus on the \emph{force-limited} case, namely when  feedback is actuated via a Hamiltonian proportional to the laboratory position operator $\hat{Q}_0$.
Remarkably, thanks to the LQG-control theory, in this case we can actually find the optimal feedback strategy, represented by a matrix $K_{\sf opt}$, with a fixed limited feedback matrix $F_{\sf m}= \mathbb{0}_2 \oplus W$, where the matrix $W$ has been defined in Eq.~\eqref{eq:matrixW}. In this case, as both the drift matrix $A$ and the feedback matrix $F_{\sf m}$ are time-dependent, one has to evaluate the time-periodic stationary matrices representing the conditional states covariance matrix $\sigmaCM_c$, the optimal Bayesian feedback matrix $Y$ and the excess noise matrix $\boldsymbol{\Sigma}_{\sf fb}$ via respectively Eqs. (\ref{eq:sigmac}), (\ref{eq:ypsilon}) and (\ref{eq:SigmaBayes}). 
The corresponding feedback Hamiltonian takes the form
\begin{align}
\mathcal{H}_{\sf fb,b} = - (\beta_Q(t) \langle \hat{Q}\rangle_c + 
\beta_X(t) \langle \hat{X}\rangle_c) \hat{Q}_0 \,,
\end{align}
where in general we observe $\beta_Q(t) > \beta_X(t) > 0$, and the extra term portional to $\langle \hat{X}\rangle_c$ arises because of the counter-rotating terms in the optomechanical Hamiltonian 

\begin{figure}[t!]
\includegraphics[width=0.99\columnwidth]{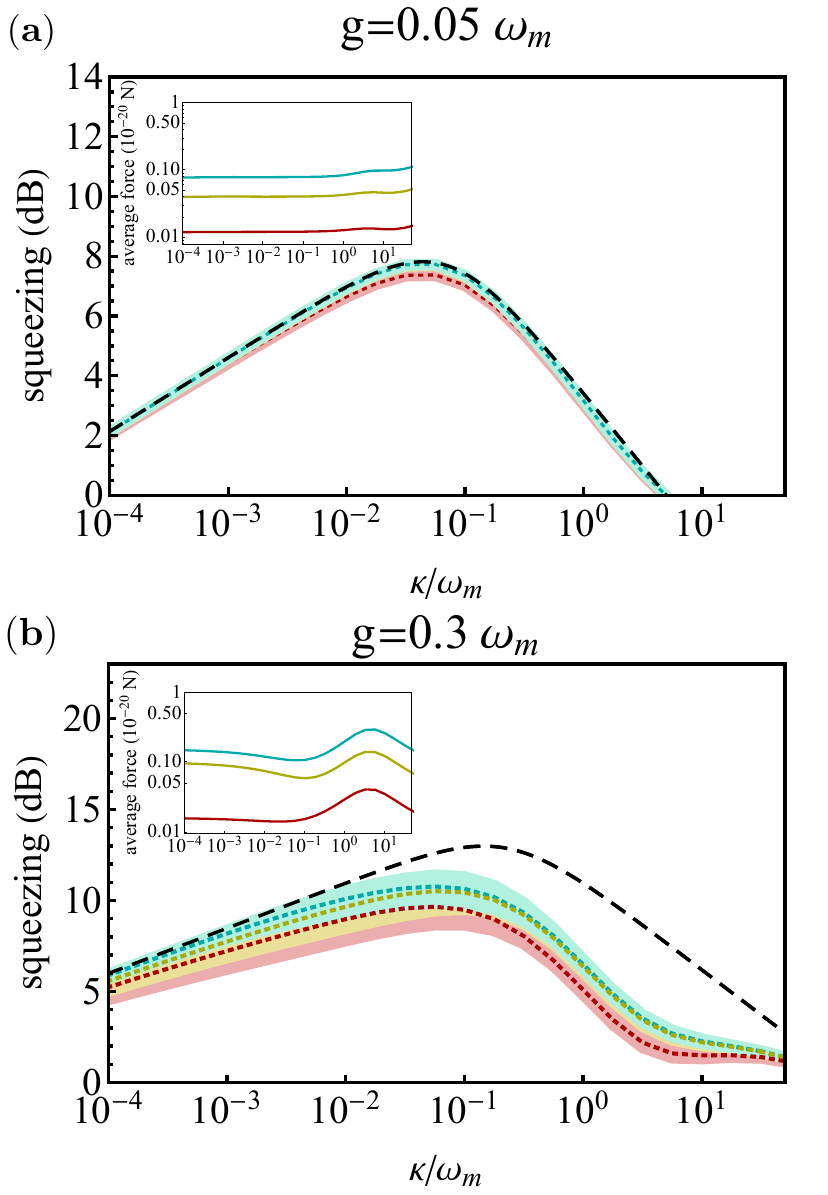}
\caption{\small{Mechanical position variance in terms of squeezing factor (dB) as a function of the sideband parameter $\kappa/\omega_{m}$, obtained unconditionally via Bayesian feedback limited to a force on the mechanical oscillator and beyond RWA for two values of the optomechanical coupling $g$ [({\bf a}), $g=0.05 \omega_m$; ({\bf b}), $g=0.3 \omega_m$] and for three different values of the cost parameter $\chi$ (from top to bottom $\chi= \{0.1, 1, 100 \}$); the dotted curves show the mean squeezing
(averaged over one mechanical period) and the shaded area extends between the minimum and maximum value of squeezing. The dashed black line denotes the optimal squeezing obtained via the conditional evolution within RWA approximation, corresponding to Eq. (\ref{RCVbrunelli}). The insets show the corresponding average feedback force $(\hbar/x_{\sf zpf}) \sqrt{\mathbbm{E}[(F{\bf u}(t))^{\sf T} F{\bf u}(t)]}$  as a function of the sideband parameter $\kappa/\omega_m$, and by considering a mechanical oscillator described by a zero point motion $x_{\sf zpf} = 10^{-14}\, m$.\\
The other parameters are fixed as following: $\eta=1$, $\bar{n}=10$, $\gamma=10^{-4}\omega_{m}$
}}\label{f:BayesianNoRWA}
\end{figure}

The corresponding unconditional fluctuations $\langle \Delta \hat{Q}^2 \rangle_{\sf fb}$ are then evaluated by averaging them over a period and we have reported them in Fig. \ref{f:BayesianNoRWA}. We remind that, at variance with Markovian feedback, here the feedback signal is  bounded. In particular one can evaluate its steady-state average magnitude $\mathbbm{E}[(F{\bf u}(t))^{\sf T} F {\bf u}(t)]$ via Eq. (\ref{eq:averageFB}). We have numerically evaluated this quantity, averaging as before its value over its steady-state period, and we have reported it in the insets of Fig. \ref{f:BayesianNoRWA} (in order to obtain the average force in Newton, we have taken its square root and multiplied it by a factor $\hbar/x_{\sf zpf}=10^{-20} N s$, with $x_{\sf zpf} \approx 10^{-14}\, m$ being the zero point motion of a standard mechanical oscillator). From these plots we can take the following conclusions: remarkably for small values of the optomechanical coupling constant $g$ (e.g. for $g=0.05\omega_m$), we obtain the optimal amount of squeezing, that is the one corresponding to Eq. (\ref{RCVbrunelli}) for conditional states evolving within the RWA approximation, even by considering the counter-rotating terms and if we restrict to a force-limited Bayesian feedback. As we increase the value of $g$ (e.g. for $g=0.3\omega_m$), we do observe sensible deviations from the optimal case, that seem to be more due to the role of counter-rotating terms, rather than to the limited feedback strategy (we remark that the results shown for $\chi =0.1\,\omega_m^{-1}$ are not much improved if we further decrease $\chi$). In particular we find that the average feedback force needed to obtain the nearly-optimal results is indeed very small. In general we observe that ${\bf u}(t)$ depends on the cost parameter $\chi$ via a constant factor, and thus, once the time-dependence of the optimal matrix $K_{\sf opt}$ has been identified, one should try to implement a corresponding feedback force ${\bf u}(t)$ proportional to the first moment vector $\bar{\bf r}_c$, and with the larger proportionality constant allowed by the experimental setup 
and our results show how nearly-optimal steady-state squeezing can be obtained by implementing a force that is well within reach of the state-of-the-art experimental capabilities.

\section{Conclusions}\label{s:Conclusions}
In this work we have discussed in detail how to \emph{deterministically} generate mechanical squeezing via time-continuous back-action evading (BAE) measurement plus feedback. 
Contrary to previous studies~\cite{CMJ2008}, our approach takes into account the effects of a finite cavity linewidth and is not limited to the weak-coupling regime; it is therefore apt to describe state-of-the-art optomechanical systems.
We discussed two main feedback strategies: state-based (Bayesian) feedback, which employs the measurement record to compute (in real-time) the optimal feedback signal and direct (Markovian) feedback, where the measured current is directly fed back to the system. In both cases, our approach consisted in first determining the form of the feedback that needs to be implemented in the case of an ideal BAE measurement; this provide a benchmark to be contrasted with realistic scenarios, where backaction evasion is imperfect and the manipulation of the optomechanical system is subject to limitations, which render feedback control non-optimal.

When employing a Markovian feedback strategy, achieving optimal unconditional squeezing requires feedback on both the cavity and the mechanical degree of freedom; only for a fast enough cavity, mechanical-only feedback recovers the maximum amount of squeezing (equal to the conditional one). Moreover, when further restricting the feedback to be actuated  only via a mechanical force, we found that counter-rotating terms cannot be neglected, due to the unbounded feedback signal. Once these are included, significant values of squeezing may be obtained for small enough coupling $g$, although force-limited feedback always adds some noise. On the other hand, Bayesian feedback attains the maximum squeezing (namely it adds no noise) via a mechanical term alone. 
Even in the case of force-feedback, for not too large couplings $g$ it attains nearly-optimal unconditional squeezing across the whole range of sideband values $\kappa/\omega_m$ and that the average feedback force needed is well within the state-of-the-art experimental capabilities. 

Therefore, we identified the conditions under which a ``cheap'' feedback, consisting of a time-dependent mechanical force (and thus easier to implement), results in no or little added noise and thus generates large amount of mechanical squeezing. We have included counter-rotating terms in our analysis, which are usually neglected. On the one hand, we found that depending on the nature of the feedback policy, the effect brought about by these terms cannot be overlooked. On the other hand, their incorporation allows to model optomechanical BAE measurements for any sideband resolution and even in the strong coupling regime.  
While in our discussion ideal detection efficiency was assumed when illustrating the results, we stress that our framework incorporates the impact non-unit efficiency of the monitoring process; this additional limitation (fundamental for experimental implementations) can be readily assessed with the expressions derived in the present work. Another source of imperfection may come from non-negligible delay time in the feedback loop. We plan to tackle delayed feedback in future works. 

Our results are directly relevant for ultra-sensitive force and displacement measurements, e.g. gravitational wave detection~\cite{Ma17}, applications to quantum information processing~\cite{Oussama15} as well as fundamental studies on the effects of quantum decoherence~\cite{Nimmrichter14,Sam17,Genoni2016Collapse}.

Our analysis could be extended to the multimode optomechanical systems consisting of two mechanical resonators coupled to a common cavity mode, for the deterministic generation of mechanical EPR entanglement and multipartite entanglement~\cite{MeBAE19,TwoModeBAE}. Finally,  besides optomechanics, our results also apply to QND measurements in hybrid quantum systems~\cite{HammererEPR,HybridBAE,Motazedifard2016,Motazedifard2019} and cavity-coupled atomic ensembles, e.g. for deterministic generation of spin squeezing~\cite{BAESpin15}.

\section{acknowledgements}
MB acknowledges support by the European Union Horizon 2020 research and innovation programme under grant agreement No 732894 (FET Proactive HOT). MGG acknowledges support by Ministero dell'Istruzione, dell'Universit\`a e della Ricerca (Rita Levi-Montalcini fellowship). 

\appendix
\section{Gaussian formalism for optomechanical Hamiltonian}\label{a:optomatrices}
In this appendix we report the matrices entering into Eqs. (\ref{eq:runc}), (\ref{eq:sigmaunc}), (\ref{eq:rc}) and (\ref{eq:sigmac}), and corresponding to the optomechanical BAE setup described in Sec.~\ref{s:BAE}. \\
The corresponding unconditional and conditional dynamics in interaction picture respect to $\mathcal{H}_0$, are described respectively by the master equation
\begin{align}
\frac{d\varrho_{\sf unc}}{dt} &= \mathcal{L}\varrho_{\sf unc}   \\
&= -i [\hat{\mathcal{H}}_{\sf int},\varrho_{\sf unc}] + \kappa\mathcal{D}[\hat{a}]\varrho_{\sf unc} 
\\
& \,\,\,\, \,\,
+ \gamma(\bar{n} + 1) \mathcal{D}[\hat{b}]\varrho_{\sf unc} + \gamma\bar{n} \mathcal{D}[\hat{b}^\dagger]\varrho_{\sf unc} \label{eq:mastereq} \,,
\end{align}
and by the stochastic master equation
\begin{align}
d\varrho_c =  \mathcal{L}\varrho_{c}\,dt + \sqrt{\eta\kappa}\mathcal{H}[-i\hat{a}]\varrho_c \,dw \,,
\end{align}
with a continuous photocurrent $I_{Y}(t)dt = -\sqrt{2 \eta \kappa}\, \langle \hat{Y} \rangle_c\,dt + dw$, and where we have defined the superoperator $\mathcal{H}[\hat{O}]\varrho = \hat{O}\varrho + \varrho \hat{O}^\dag - \Tr[\varrho (\hat{O} + \hat{O}^\dag)]\varrho$. 
The matrices $A$, $D$, $E$, and $B$ can be derived by following the formalism introduced in \cite{AlessioBook,Marco16}, or analogously in \cite{WisemanDoherty}. We start by presenting the scenario where the RWA can be performed, and the optomechanical Hamiltonian reads
\begin{align}
\mathcal{H}_{\sf int} = -g \hat{X} \hat{Q} \,, \nonumber
\end{align}
The corresponding matrices are
\begin{align}
A &=
\left( 
\begin{array}{c c c c}
-\frac{\kappa}{2} & 0 & 0 & 0 \\
0 & -\frac{\kappa}{2} & g & 0 \\
0 & 0 & -\frac{\gamma}{2} & 0 \\
g & 0 & 0 & -\frac{\gamma}{2}
\end{array}
\right), \nonumber \\
D &=
\left( 
\begin{array}{c c c c}
\kappa & 0 & 0 & 0 \\
0 & \kappa & 0 & 0 \\
0 & 0 & \gamma(2 \bar{n} +1) & 0 \\
0 & 0 & 0 &  \gamma(2 \bar{n} +1)
\end{array}
\right), \nonumber \\
B &= E = 
\left( 
\begin{array}{c c c c}
0 & 0 & 0 & 0 \\
0 & \sqrt{\eta\kappa} & 0 & 0 \\
0 & 0 & 0 & 0 \\
0 & 0 & 0 & 0 
\end{array}
\right).\nonumber 
\end{align}
In the regime where the RWA does not apply and thus the optomechanical Hamiltonian is described by Eq. (\ref{optom2tone}),
\begin{align}
\hat{\mathcal H}_{\sf int}(t)=-g\hat{X}\left[  \hat{Q}\left( 1+\cos(2\omega_{m}t)\right) +\hat{P}\sin(2\omega_{m}t)\right], \nonumber
\end{align}
the drift matrix $A$ is the only matrix that changes its form, it becomes time-dependent and it reads
\begin{widetext}
\begin{align}
A &=
\left( 
\begin{array}{c c c c}
-\frac{\kappa}{2} & 0 & 0 & 0 \\
0 & -\frac{\kappa}{2} & g (1 + \cos(2\omega_m t)) & g \sin(2 \omega_m t) \\
-g \sin(2 \omega_m t) & 0 & -\frac{\gamma}{2} & 0 \\
g(1+\cos(2\omega_m t)) & 0 & 0 & -\frac{\gamma}{2}
\end{array}
\right). \nonumber 
\end{align}
\end{widetext}
\section{Derivation of Lyapunov equation for the excess noise matrix $\boldsymbol{\Sigma}$}\label{a:AppendixB}
Let us consider the following stochastic evolution of the first moment vector
\begin{align}
d\bar{\bf r}_c = \tilde{A}\bar{\bf r}_c dt + V \frac{d{\bf w}}{\sqrt{2}} \,. \label{eq:rcgeneral}  
\end{align}
Our goal is to derive the evolution equation for the excess noise matrix $\Sigma$, defined in Eq. (\ref{eq:excessnoise}) as 
\be
\boldsymbol{\Sigma} =  \EE[ \{\bar{\bf r}_c ,\bar{\bf r}_c^{\sf T} \} ] -  \{ \EE[\bar{\bf r}_c] , \EE[\bar{\bf r}_c^{\sf T} ] \}.
\ee 
By deriving the first term respect to time, and by exploiting Ito calculus, one obtains
\begin{widetext}
\begin{align}
d (  \EE[ \{\bar{\bf r}_c ,\bar{\bf r}_c^{\sf T} \} ]) &= 
 \EE[ \{d\bar{\bf r}_c ,\bar{\bf r}_c^{\sf T} \} ] +  \EE[ \{\bar{\bf r}_c ,d\bar{\bf r}_c^{\sf T} \} ] +  \EE[ \{d\bar{\bf r}_c ,d\bar{\bf r}_c^{\sf T} \} ] 
\nonumber \\
&=  \tilde{A} \, \EE[ \{\bar{\bf r}_c ,\bar{\bf r}_c^{\sf T} \} ] \,dt + \EE[ \{\bar{\bf r}_c ,\bar{\bf r}_c^{\sf T} \} ] \tilde{A} \,dt \nonumber + V \left( \mathbbm{E}\left[\frac{ \{d{\bf w}, d{\bf w}^{\sf T} \}}{2} \right] \right) V^{\sf T}  \nonumber  \\
&=  \left( \tilde{A} \, \EE[ \{\bar{\bf r}_c ,\bar{\bf r}_c^{\sf T} \} ]  + \EE[ \{\bar{\bf r}_c ,\bar{\bf r}_c^{\sf T} \} ] \tilde{A}^{\sf T} + V V^{\sf T} \right) dt\,, \nonumber
\end{align}
\end{widetext}
where we have exploited the Wiener increments property $\{d{\bf w}, d{\bf w}^{\sf T} \}/2 = \mathbbm{1}\,dt$.\\
The second term on the other hand yields 
\begin{widetext}
\begin{align}
d\left(\{ \EE[\bar{\bf r}_c] , \EE[\bar{\bf r}_c^{\sf T} ] \}\right) &=  \{ \EE[d\bar{\bf r}_c] , \EE[\bar{\bf r}_c^{\sf T} ] \} +  \{ \EE[\bar{\bf r}_c] , \EE[d\bar{\bf r}_c^{\sf T} ] \} \nonumber\\
&= \left(\tilde{A}  \{ \EE[\bar{\bf r}_c] , \EE[\bar{\bf r}_c^{\sf T} ] \}  +  \{ \EE[\bar{\bf r}_c] , \EE[\bar{\bf r}_c^{\sf T} ] \} \tilde{A}^{\sf T} \right) dt \,. \nonumber
\end{align}
\end{widetext}
By combining the two equations, we finally find the Lyapunov equation for the excess noise matrix
\begin{align}
\frac{d\boldsymbol{\Sigma}}{dt} = \tilde{A}\boldsymbol{\Sigma} + \boldsymbol{\Sigma}\tilde{A}^{\sf T} + V V^{\sf T} \,.
\end{align}

As a consequence, given the evolution of the first moment vector $\bar{\bf r}_c$ ruled by either Eq. (\ref{eq:rcFBMarkov}) for Markovian feedback, or Eq. (\ref{eq:rcFBBayes}) for Bayesian feedback, 
one finds that the evolution for the excess noise matrix 
$\boldsymbol{\Sigma}$ is given respectively by Eq. (\ref{eq:SigmaMarkov}) or Eq. (\ref{eq:SigmaBayes}).
\bibliography{MyRefs}
\end{document}